# Search for the Decay $K^+ \to \pi^+ \nu \bar{\nu}$

S. Adler, M. S. Atiya, I-H. Chiang, J. S. Frank, J. S. Haggerty, T. F. Kycia, K. K. Li,
L. S. Littenberg, A. Sambamurti,[a] A. Stevens, R. C. Strand, and C. Witzig

*Brookhaven National Laboratory, Upton, New York 11973*

W. C. Louis

*Medium Energy Physics Division, Los Alamos National Laboratory, Los Alamos, New Mexico 87545*

D. S. Akerib,[b] M. Ardebili, M. R. Convery, M. M. Ito, D. R. Marlow, R. A. McPherson,
P. D. Meyers, M. A. Selen,[c] F. C. Shoemaker, and A. J. S. Smith

*Joseph Henry Laboratories, Princeton University, Princeton, New Jersey 08544*

E. W. Blackmore, D. A. Bryman, L. Felawka, P. Kitching, A. Konaka, V. A. Kujala,
Y. Kuno,[d] J. A. Macdonald, T. Nakano,[e] T. Numao, P. Padley,[f] J-M. Poutissou,
R. Poutissou, J. Roy,[g] R. Soluk, and A. S. Turcot[h]

*TRIUMF, Vancouver, British Columbia, Canada V6T 2A3*

(17 October 1995)

## Abstract

An upper limit on the branching ratio for the decay $K^+ \to \pi^+ \nu \bar{\nu}$ is set at $2.4 \times 10^{-9}$ at the 90% C.L. using pions in the kinematic region 214 MeV/c < $P_\pi$ < 231 MeV/c. An upper limit of $5.2 \times 10^{-10}$ is found on the branching ratio for decays $K^+ \to \pi^+ X^0$, where $X^0$ is any massless, weakly interacting neutral particle. Limits are also set for cases where $M_{X^0} > 0$.

The process $K^+ \to \pi^+ \nu \overline{\nu}$ is highly suppressed in the Standard Model by the Glashow-Iliopoulos-Maiani (GIM) mechanism [1], which forbids flavor-changing neutral currents (FCNC) at tree level. The unequal masses of the virtual up, charm, and top quarks slightly spoil the cancellation in one-loop diagrams, allowing a small amplitude to survive. The large top quark mass enhances this effect, making the virtual top quark the largest contributor to $K^+ \to \pi^+ \nu \overline{\nu}$ in the Standard Model. Although the charm quark contribution is not negligible, recent theoretical calculations [2] have greatly reduced the uncertainty from virtual charm quark lines making a measurement of the $K^+ \to \pi^+ \nu \overline{\nu}$ branching ratio an extremely clean measurement of the product of the Cabibbo-Kobayashi-Maskawa quark-mixing matrix ($V_{\mathrm{CKM}}$ [3]) elements $V_{ts}^* V_{td}$. Currently, this product is constrained primarily by measurements of $V_{cb}$ and $|V_{ub}/V_{cb}|$ through the mixing matrix unitarity [4], but is also further restricted by measurements of other higher-order processes, though with significant theoretical uncertainty. Combining the recent $K^+ \to \pi^+ \nu \overline{\nu}$ calculations from Reference [2] with constraints on $V_{\mathrm{CKM}}$ parameters from CP violation in the neutral kaon system and $B^0 - \overline{B}^0$ mixing [5], and using the current measurements of the top quark mass [6], the $K^+ \to \pi^+ \nu \overline{\nu}$ branching ratio is most likely in the range $(0.6 - 3) \times 10^{-10}$. New processes such as direct FCNC or a decay $K^+ \to \pi^+ X^0 X^0$, where $X^0$ is a new weakly interacting neutral particle, could give a three-body branching ratio outside of this range. A two-body decay $K^+ \to \pi^+ X^0$ would also be new physics, with a cleaner experimental signature. Proposed $X^0$ candidates are supersymmetric particles [7] or light Goldstone bosons (e.g., Majorons [8], familons [9], or axions [10]).

The E787 detector [11] operating with the low energy separated beam (LESB I) at the AGS at Brookhaven National Laboratory collected data for a total of seven months during the 1989, 1990 and 1991 running periods. We previously reported 90% confidence level limits [12] on the branching ratios, $B(K^+ \to \pi^+ \nu \overline{\nu}) < 5.2 \times 10^{-9}$ and $B(K^+ \to \pi^+ X^0) < 1.7 \times 10^{-9}$ ($M_{X^0} = 0$) based on data from the 1989 run alone. Here, we add those data to similar sets from the 1990 and 1991 runs and analyze the combined set. The 1989-1991 sample is the complete data set taken before major upgrades to the beamline and detector were undertaken.

The experiment can be summarized as follows: 800 MeV/$c$ kaons, identified by Čerenkov and $dE/dX$ counters, are stopped in a segmented active target (see Figure 1), where they decay. Independent measurements of the kinetic energy, momentum, and range (in scintillator) of the charged decay products are made using the target, the central drift chamber, and the cylindrical range stack of plastic scintillator layers. Pions are distinguished from muons kinematically and by identifying the $\pi \to \mu \to e$ decay sequence using 500-MHz transient digitizers. Photons are detected in a nearly $4\pi$ solid-angle lead-scintillator calorimeter that is 12–14 radiation lengths thick. The entire detector is in a 1 T solenoidal magnetic field for the momentum measurement.

The signature for $K^+ \to \pi^+ \nu \overline{\nu}$ is a $K^+$ decay with a $\pi^+$ as the only observable product. The two dominant $K^+$ decay modes are the most important background sources. $K^+ \to \mu^+ \nu_\mu$ ($K_{\mu2}$), a two-body decay with a 64% branching ratio, produces a 236 MeV/$c$ $\mu^+$. $K^+ \to \pi^+ \pi^0$ ($K_{\pi2}$), a two-body decay with a 21% branching ratio, produces a 205 MeV/$c$ $\pi^+$. Since the $K^+ \to \pi^+ \nu \overline{\nu}$ momentum spectrum extends to 227 MeV/$c$, we can search for it either above or below the $K_{\pi2}$ peak. While there is more phase space with $P_\pi < 205$ MeV/$c$, interactions with detector material can shift a $K_{\pi 2}$ pion down into this region making the background severe [13]; in this paper, we report on the search above the $K_{\pi 2}$ peak. The dominant source of pions with $P_\pi > 205$ MeV/c is beam pions — about two-thirds of the beam particles are pions — that scatter from the target into the range stack.

The search for $K^+ \to \pi^+ \nu \overline{\nu}$ follows a threefold strategy: (i) the incident beam particle is identified as a $K^+$ that has stopped in the target, with no beam particles at the apparent kaon decay time; (ii) the only observed decay product is a single charged-particle track identified as a $\pi^+$ that must be delayed in time with respect to the kaon; (iii) the energy, range, and momentum of the $\pi^+$ each lie between the $K_{\pi 2}$ and $K_{\mu 2}$ peaks. A multilevel trigger employs each of these elements to reject background events online, while the analysis makes more refined use of detector information for further rejection. $K_{\mu 2}$ can survive only if the muon is misidentified as a pion and the kinematics are reconstructed incorrectly. $K_{\pi 2}$ can survive only if both photons from the $\pi^0$ decay are missed and the kinematics are reconstructed incorrectly. Scattered beam pions can survive only if the $\pi^+$ is misidentified as a $K^+$ with the scattered track mismeasured to be delayed, or if it is missed by the beam counters and follows a $K^+$. The measures taken to deal with the main background sources are also very effective against other backgrounds, such as radiative $K_{\mu 2}$ decays or $K^+$ charge exchange interactions followed by $K^0_L \to \pi^+ l^- \overline{\nu}_l$. Backgrounds from other $K^+$ decay modes were examined and found to be negligible.

After establishing the overall analysis strategy, we adjusted the cuts with the intention of reducing the total expected background to significantly under one event in the final sample. The final cuts used were developed during studies of the known background processes. In these studies, we take advantage of the redundant methods available for the rejection of each background by dividing the cuts used to suppress it into two groups. One group of cuts is relaxed or inverted to enhance the background sample, then the other group is applied and its rejection is measured. This background-study technique allows us to use data to infer background levels of less than one event. For example, a large sample of $K_{\mu 2}$ background events is obtained by removing the transient digitizer particle identification cuts, and this sample is used to measure the $K_{\mu 2}$ rejection of the kinematic analysis. Similarly, the transient digitizer rejection is measured with kinematically selected muons. Assuming these rejections are independent, they are combined and used to estimate the number of $K_{\mu 2}$ events that will survive the full analysis. Correlations between the two groups of cuts will introduce an error in the background estimates from this method, and we group the cuts to minimize these effects.

The detector calibration procedures and analysis software used for the final analysis presented here [14] have been refined considerably since Reference [12]. Improvements included increased acceptance of the transient digitizer particle identification cuts and improved kinematic resolutions with reduced kinematic tails. An initial analysis [15] had been completed before the calibration and software improvements were finished, observing background in excess of predictions. The final analysis had a significantly higher acceptance and did not suffer from some anomalies in the transient digitizer signals and the kinematic reconstruction that may have affected the background predictions in the initial analysis.

The background from $K_{\mu 2}$ (including $K^+ \to \mu^+ \nu_\mu \gamma$) was evaluated by separately measuring the rejections of the transient digitizer particle identification and kinematic cuts, and is estimated to be less than 0.15 events. The background from $K_{\pi 2}$ was evaluated by sep-arately measuring the rejections of the photon veto and kinematic cuts, and is estimated to be less than 0.14 events. The background from beam pion scattering was evaluated by separately measuring the rejections of the beam counter and timing cuts, and is estimated to be less than 0.07 events. Monte Carlo studies indicated that the background from $K^+$ charge exchange interactions was about 0.1 events.

Figure 2(a) shows the range in scintillator versus kinetic energy for charged tracks in the final sample. Only events with a measured charged track momentum in the accepted region $211 \leq P_\pi \leq 243$ MeV/c are plotted. The rectangular box defines the search region in kinetic energy ($115 \leq T_\pi \leq 135$ MeV, corresponding to $213 \leq P_\pi \leq 236$ MeV/c) and range ($34 \leq R_\pi \leq 40$ cm of scintillator, corresponding to $214 \leq P_\pi \leq 231$ MeV/c), and encloses the upper 15% of the $K^+ \to \pi^+ \nu \overline{\nu}$ phase space. Note that the range cut defines the kinematic search region. No events are observed in the signal region. There are six events above the signal region, which is consistent with the $5.5 \pm 0.6$ expected from the $K_{\mu 2}$ background study. The events clustered at $T_\pi = 108$ MeV and $R_\pi = 30$ cm are $K_{\pi 2}$ decays where both photons from the $\pi^0$ are missed. The number of such events is consistent with Monte Carlo estimates of the photon detection inefficiency [16].

Where possible, we used calibration data taken simultaneously with the physics data for the acceptance calculation. We relied on Monte Carlo for only the solid angle coverage, the accepted region of the $\pi^+$ spectrum, and the losses from $\pi^+$ nuclear interactions and decays in flight. The $\pi^+$ spectrum for $K^+ \to \pi^+ \nu \overline{\nu}$ was calculated using a Standard Model matrix element with massless neutrinos [17]. Figure 2(b) shows the spectrum of Monte-Carlo-simulated $K^+ \to \pi^+ \nu \overline{\nu}$ after the full analysis. $K_{\mu 2}$ calibration data were used to measure losses from the beam analysis, from the charged track reconstruction inefficiency, from the $K^+ \to \pi^+$ delayed coincidence requirement, and from accidental energy depositions at the kaon decay time above our approximately 1 MeV photon veto threshold. Scattered beam pion data were used to measure the acceptance of the transient digitizer and kinematic $\pi^+/\mu^+$ separation cuts. The acceptance calculation is summarized in Table I, resulting in a total acceptance of 0.0027 for $K^+ \to \pi^+ \nu \overline{\nu}$ and 0.0127 for $K^+ \to \pi^+ X^0$ ($M_{X^0} = 0$). The uncertainty in the acceptance has a negligible effect on limits set with these data.

During typical running conditions, $3 \times 10^5$ kaons entered the stopping target per 1.5 s beam spill. We measure the fraction that decayed at rest in the target to be 0.65 using an analysis of $K_{\mu 2}$ data and the well known $K_{\mu 2}$ branching ratio. This normalization to $K_{\mu 2}$ removes some sources of systematic error from our sensitivity. Our final measured exposure for these data is $3.49 \times 10^{11}$ stopped kaons. The acceptance (especially the Monte Carlo simulation of $\pi^+$ nuclear interactions) and stopping fraction are checked by measuring the $K_{\pi 2}$ branching ratio, for which we obtain $0.205 \pm 0.006$, where the error includes counting statistics and other items not common to the $K^+ \to \pi^+ \nu \overline{\nu}$ analysis. This is consistent with the world average, $0.2116 \pm 0.0014$ [4].

Since no events were observed in the signal region (Figure 2(a)), we obtain 90% confidence level upper limits of $B(K^+ \to \pi^+ \nu \overline{\nu}) < 2.4 \times 10^{-9}$ [18] and $B(K^+ \to \pi^+ X^0) < 5.2 \times 10^{-10}$ ($M_{X^0} = 0$) for these data. Figure 3 shows the limit as a function of $M_{X^0}$ for different lifetime assumptions, assuming that the photon veto rejects events where the $X^0$ decays in the spectrometer.

We gratefully acknowledge the dedicated efforts of the technical staffs supporting this experiment and of the Brookhaven AGS Department. This research was supported in part

FIGURES

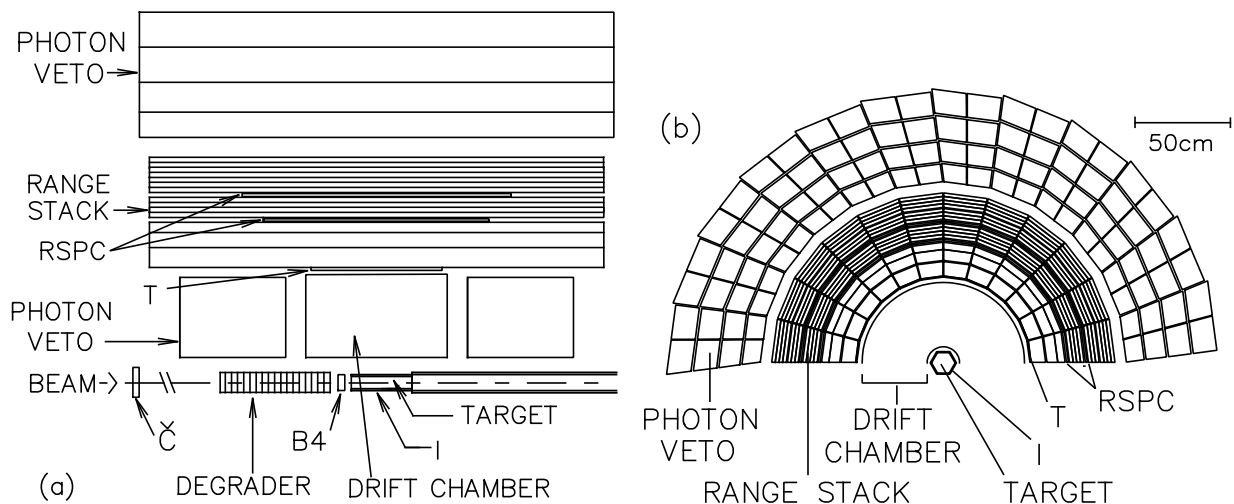

FIG. 1. Schematic (a) side and (b) end views showing the upper half of the E787 detector. Č: beam Čerenkov counter; B4: beam hodoscope; I and T: trigger scintillators; RSPC: multiwire proportional chambers.

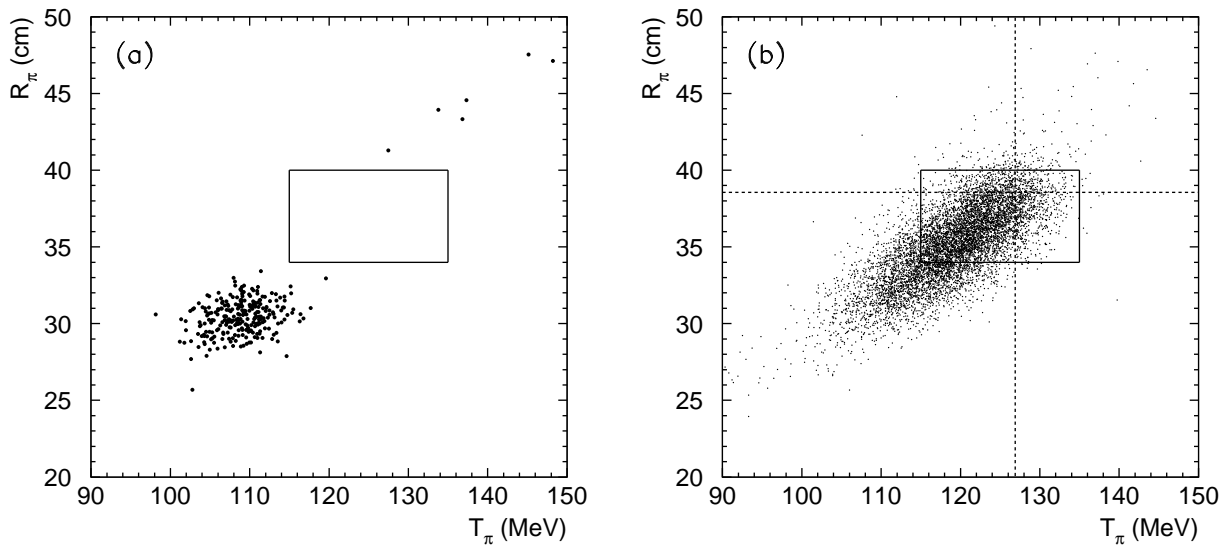

FIG. 2. Charged-track range vs. kinetic energy for (a) data and (b) $K^+ \to \pi^+ \nu \bar{\nu}$ Monte Carlo for events satisfying the selection criteria (see text) and having measured momentum $211 \le P_\pi \le 243$ MeV/c. The rectangular box indicates the search region for $K^+ \to \pi^+ \nu \bar{\nu}$ and $K^+ \to \pi^+ X^0$ ($M_{X^0} \approx 0$). The horizontal and vertical dashed lines in (b) are the theoretical end-points of $K^+ \to \pi^+ \nu \bar{\nu}$ in range and energy respectively.

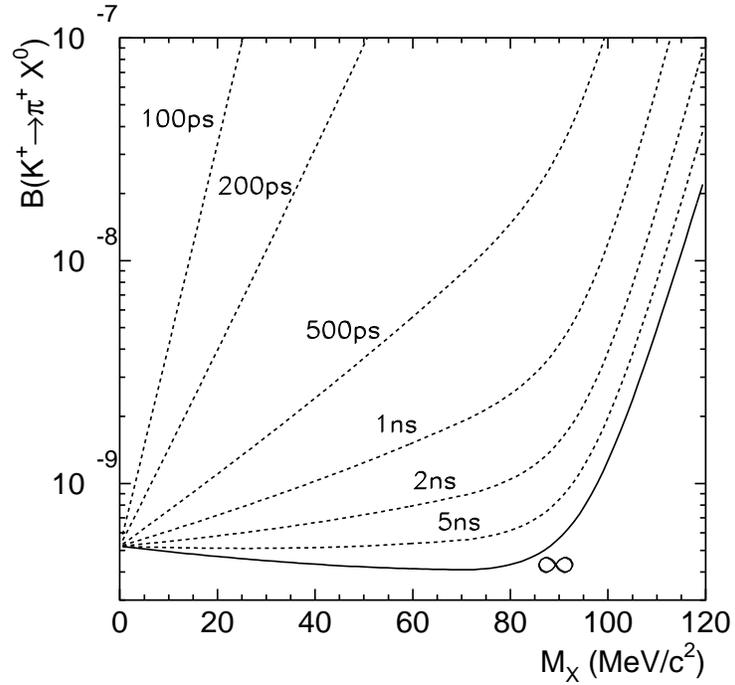

FIG. 3. The solid curve gives the 90% confidence level upper limit on the branching ratio for $K^+ \to \pi^+ X^0$ as a function of $M_{X^0}$. The dashed curves give 90% confidence level upper limits for cases where $X^0$ has a finite lifetime.

TABLES

TABLE I. Acceptance factors for $K^+ \to \pi^+ \nu \overline{\nu}$ and $K^+ \to \pi^+ X^0$ ($M_{X^0} = 0$). Each table entry represents the acceptance from a number of related cuts.

| Category | $\pi^+ \nu \overline{\nu}$ | $\pi^+ X^0$ |
|---|---|---|
| Solid angle | 0.43 | 0.43 |
| $\pi^+$ spectrum | 0.15 | 0.73 |
| $\pi^+$ nuclear absorption | 0.53 | 0.50 |
| $\pi^+$ decay in flight | 0.92 | 0.92 |
| $K^+ \to \pi^+$ delayed coincidence | 0.75 | 0.75 |
| $\pi^+/\mu^+$ kinematics | 0.87 | 0.88 |
| $\pi^+ \to \mu^+$ transient digitizer tagging | 0.41 | 0.41 |
| $\mu^+ \to e^+$ transient digitizer tagging | 0.84 | 0.84 |
| Accidental vetoes | 0.67 | 0.67 |
| Beam analysis | 0.84 | 0.84 |
| Reconstruction | 0.69 | 0.69 |
| Net Acceptance | 0.0027 | 0.0127 |